\documentclass[twocolumn,pra,aps,showpacs]{revtex4}
\def\be{\begin{equation}}
\def\ee{\end{equation}}
\def\ba{\begin{eqnarray}}
\def\ea{\end{eqnarray}}
\def\half{{1 \over 2}}
\def\nave{n}
\def\dn{\delta n}
\def\dtau{\Delta \tau}
\def\dx{\Delta x}
\def\dt{\Delta t}

\def\k{K}
\def\X{\Delta X}
\def\x{x}

\def\rate{{\cal R}}

\def\Tr{{\rm Tr}}
\def\ttheta{\widetilde{\theta}}
\def\tnu{\widetilde{\nu}}
\def\ok{\omega_k}
\def\rg{\sqrt{g}}
\def\per{{\cal T}}
\def\tW{\widehat{W}}
\input epsf
\begin{document}
\title{Tunneling in a uniform one-dimensional superfluid: emergence of a complex 
instanton}
\author{S. Khlebnikov}
\affiliation{Department of Physics, Purdue University, West Lafayette, 
IN 47907, USA}
\begin{abstract}
In a uniform ring-shaped one-dimensional superfluid, quantum fluctuations that 
unwind the order parameter need to transfer momentum to quasiparticles (phonons). 
We present a detailed calculation of the leading exponential factor governing the 
rate of such phonon-assisted tunneling in a weakly-coupled Bose gas at a low 
temperature $T$. We also estimate the preexponent. We find that for small
superfluid velocities the $T$-dependence 
of the rate is given mainly by $\exp(-c_s P/ 2T)$, where $P$ is the momentum 
transfer, and $c_s$ is the phonon speed. At low $T$, this represents a strong 
suppression of the rate, compared to the non-uniform case.
As a part of our calculation, we identify a complex instanton, 
whose analytical continuation to suitable real-time segments is real and describes 
formation and decay of coherent quasiparticle states with nonzero total momenta.
\end{abstract}
\maketitle
\section{Introduction}
Many one-dimensional (1D) systems share a universal
low-energy description based on a complex order parameter \cite{Haldane}.
Probably the most familiar example is a superfluid confined
to a narrow channel, but---due to the well-known ``duality'' between 
bosons and fermions in one dimension---a similar description exists also for 
fermionic fluids.

In accordance with the Bogoliubov-Hohenberg theorem, no long-range order
is possible in these 1D systems, but the precise nature of fluctuations that
prevent ordering deserves a further discussion. 
At zero temperature ($T=0$), perturbative fluctuations of the phase of 
the order parameter
(phonons in a superfluid) cause a power-law decay of spatial correlations.
At $T\neq 0$, the decay becomes exponential. However, a detailed study
\cite{Haldane} of the $T=0$ case reveals additional contributions 
to the correlation functions, of the form of a power law multiplied
by an oscillating factor. 

While for weakly-coupled Fermi systems these oscillating terms
can be seen as a 1D version of Friedel oscillations, for weakly-interacting
bosons their interpretation is not immediately obvious. 
It is possible, however, to interpret them
as a consequence of nonperturbative fluctuations: instantons or quantum phase
slips (QPS). The oscillatory dependence on the spatial coordinate can be traced 
to the fact (readily verified, see below) 
that each QPS changes the linear momentum of the superfluid component.

The momentum production by QPS results from unwinding the order parameter and 
the corresponding change in the supercurrent. More formally, it can be viewed as
a consequence of a special type of
topological term, present in 
the action of a 1D weakly-coupled Bose gas. We discuss this term in detail in the
next section. 

A complementary picture is obtained by looking at the Lieb-Liniger spectrum 
\cite{Lieb&Liniger} (for a gas with a delta-function repulsion). 
Their results apply for periodic boundary conditions, i.e., ring geometry,
which is the only case we consider here.
In the limit $L\to \infty$, where $L$ is 
the length of the ring, the spectral branch associated with solitons
\cite{solitons} touches zero at momentum $P = 2\pi n$, where $n$ is the gas density. 
This state corresponds precisely to the order parameter winding once as we go 
around the ring. Phonon-assisted
transitions to this state will destroy superfluidity. Our aim will be to calculate
the rate of such transitions at low temperatures.

Except for a brief summary in Sect. \ref{sect:T=0} of results for QPS induced by
a localized perturbation, we consider here only uniform 1D superfluids. 
In the ring geometry, momentum in the longitudinal
($x$) direction is conserved. (More precisely, we should be talking about angular 
momentum, but this distinction will not be important for our purposes.) Our goal 
was to see how momentum conservation influences the QPS rate. Some results of this
work have been presented in ref. \cite{slip}. Here we describe a different, more
systematic method, which confirms the results of \cite{slip}, but also allows us
to obtain new results.

At $T=0$, the 1D gas can be mapped on a two-dimensional (2D) model by introducing 
the imaginary (Euclidean) time $\tau$. QPS are vortices---or instantons---of this
2D model. Although this model is similar to the 
usual XY model, the topological term drives it into a different universality class.
The XY model, as the coupling is increased, 
undergoes a Berezinsky-Kosterlitz-Thouless (BKT) transition. In contrast, in the
Bose gas, the correlation functions \cite{Haldane} evolve 
continuously. We show in Sect. \ref{sect:T=0} that, for a weakly-coupled uniform 
Bose gas at zero temperature, instantons and antiinstantons are bound in pairs by 
a linear, rather than logarithmic, potential.
An extrapolation of this result to strong coupling 
implies that the breaking of instanton pairs, 
characteristic of a BKT transition, is not possible, a conclusion consistent with the 
expected Galilean invariance of the $T=0$ state. 

Instanton-antiinstanton pairs can unbind if there is an additional source of energy,
besides the energy resulting from unwinding the supercurrent. This possibility may
be of interest for analog models of gravity \cite{Unruh,Garay&al}.
It has been suggested that one can use cold Bose gases to model 
cosmologically interesting spacetimes
\cite{Fedichev&al,Barcelo&al}. If one models an expanding spacetime by
varying (decreasing) the coupling $g$ between the atoms, as proposed in 
ref. \cite{Barcelo&al}, then
some of the energy supplied by this variation may become available 
for enhancement of QPS. We note also that a decrease in $g$ causes the principal
length scale of the gas---the ``healing'' length $\xi$---to grow. 
Since the cross-sectional radius $R$ of the channel is fixed,
the 1D regime $\xi > R$ can be reached.

Another possibility, which is the main subject of this paper, is when 
the additional source of energy is a low but nonzero temperature.
We discuss this case in detail in Sect. \ref{sect:Tneq0}. 
Thermally activated phase slips (TAPS) in the ring geometry 
have been considered, in the framework of the nucleation theory of
Refs. \cite{Little,LA,McCH}, in ref. \cite{TAPS}. 
In the present paper, we concentrate on lower
temperatures where, as we will see, the main mechanism for phase slips is
thermally assisted quantum tunneling, rather than the over-barrier activation.
One should note, however, that in the case of a {\em uniform} Bose gas, even
at higher temperatures, 
the original LAMH theory \cite{LA,McCH} probably needs to be modified,
in order to account for momentum conservation. We
discuss this point further in the conclusion.

If the gas is uniform (or nearly uniform---the confining potential is smooth on
the scale $\xi$, and its variations are small),
QPS have a rather distinctive experimental signature. Indeed, as we will
see, unwinding momentum $P$ from the supercurrent in such a uniform system
is accompanied by transferring a compensating momentum to the phonon
``bath''. This is equivalent to creating a flow of excited atoms, which can in 
principle be detected experimentally. For example, we can start with a state with
no supercurrent and let a QPS create a unit of supercurrent and a compensating
normal flow in the opposite direction (counterflow). The counterflow can be
detected by the standard momentum imaging, i.e., opening the trap in one place. 
Note that no counterflow is expected for QPS induced by a
highly localized perturbation,
which breaks both momentum conservation and the Galilean invariance (calculation
of the rate for this case has been done in ref. \cite{Kagan&al}). In that case, 
the final state of phonons has zero total momentum.

However, perhaps the most immediate experimental consequence of the momentum balance
during QPS in a uniform system is that it leads to a strong suppression of the rate,
compared to the case of a localized perturbation \cite{slip}. The reason is that
the momentum released by unwinding the supercurrent can only be absorbed by relatively
high-energy phonons states, which are scarcely populated at a low temperature.

From the theoretical perspective, one would like to understand in general 
how to compute the
rates of instanton processes that transfer momentum between the background 
and excitations.
The question is not limited to QPS in narrow superfluid channels but arises also
in other contexts. For example, one can view the momentum transfer by QPS
as a 1D analog of the Magnus force in higher dimensions. 
This force acts on vortices moving
in a superfluid and is known to suppress vortex tunneling \cite{Volovik}. 
It is natural to ask if the
suppression can be circumvented by an inelastic process similar to the one we 
consider here. Furthermore, additional interesting physics \cite{slip}
emerges in cases when the order parameter is coupled to fermions,
as in the case of BCS superconductors. The coupling to fermions opens a new
channel of momentum production, due to
the fermion zero modes at the instanton core. This channel is a 1D analog
of ``momentogenesis'' \cite{momen} by vortices in 2D arrays of Josephson junctions.
Although in the case of 1D superconductors the relevance of this channel
is somewhat obscured by scattering of quasiparticles on the boundaries
and disorder, it may still be of interest for interpretation
of experiments \cite{exp1,exp2} on superconducting nanowires.

A much studied example of inelastic tunneling in field theory is instanton-induced
scattering in gauge theories and their low-dimensional analogs \cite{instantons}.
Indeed, the complex instanton that we will find in this paper is a generalization
of the (real) periodic instantons \cite{periodic} to the case when there is 
nontrivial momentum transfer from the background to quasiparticles.

\begin{figure}
\leavevmode\epsfxsize=3.25in \epsfbox{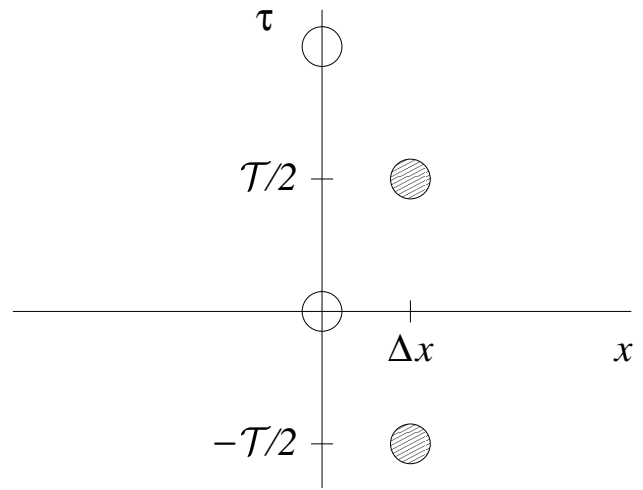}
\vspace*{0.2in}
\caption{A periodic configuration, in which instantons (open circles) and 
antiinstantons (shaded circles) are shifted relative to each other 
in space by $\dx$.
At a finite temperature $T$, the period of the configuration is 
$\per = \beta = 1/T$.}
\label{fig:inst}
\end{figure}

In ref. \cite{slip}, we
have identified periodic Euclidean configurations 
(of period $\beta=1/T$, where $T$ is the
temperature), which consist of chains of instantons and antiinstantons shifted
relative to each other by amount $\beta/2$ in the imaginary time and by some $\dx$
in space, see Fig. \ref{fig:inst}. 
The tunneling rate has been obtained by integrating over $\dx$. 
Unlike the case without momentum transfer  \cite{periodic}, this integration is 
nontrivial. Nevertheless, the leading exponential factor obtained in this way
has a simple physical interpretation. It can be interpreted as
the rate of tunneling between quasiparticle (phonon) states
with momenta $-P/2$ and $P/2$, so that the change in the momentum of phonons
precisely compensates the momentum produced by unwinding the supercurrent.

In the present paper, we derive this leading result for the tunneling exponent, 
and the first correction to it, in what we regard as a more systematic 
way. First, using a method developed in ref. \cite{periodic},
we obtain the tunneling exponent for a microcanonical state (fixed energy $E$).
Then, we integrate the microcanonical rate over $E$ with
the Boltzmann factor $\exp(- E/T)$ to obtain the canonical (fixed $T$) rate.
This method allows us to find directly the energies 
corresponding to the dominant phase-slip paths. In particular, we can show that 
at sufficiently low $T$ thermally-assisted tunneling (as opposed to over-barrier
activation) is indeed the dominant phase-slip mechanism.

We also discuss in detail the instanton solution that saturates the rate at nonzero
momentum transfer $P$. This instanton corresponds to a 
complex saddle point for $\dx$ and is itself complex. Unlike the periodic
instanton of ref. \cite{periodic}, it has no turning points. Nevertheless,
it is possible to identify the initial and final states connected by this complex
instanton and to reconstruct their real-time evolution. 
On the real-time segments, the solution is
real and can be interpreted as formation and decay of coherent phonon states
corresponding to the tunneling endpoints. A direct computation confirms that 
the total momenta of these states are $-P/2$ and $P/2$.

\section{The topological term}
\label{sect:top_term}
A weakly-coupled 1D Bose gas can be described by the Gross-Pitaevsky (GP) Lagrangian,
which in the Euclidean signature reads
\be
L_E =  \psi^{\dagger} \partial_\tau \psi + \frac{1}{2m} |\partial_x \psi|^2 
+ {g \over 2} |\psi|^4 - \mu |\psi|^2 \; .
\label{L1}
\ee
Here $\tau= it$ is the Euclidean time, $m$ is the mass of the particles, $g >0$ 
is the coupling constant, and $\mu$ is the chemical potential. We assume that the
system is subject to periodic boundary conditions in the
$x$ direction: $\psi(x+L) = \psi(x)$. 

Instantons are vortices of this theory in the
$(x,\tau)$ plane, corresponding to nontrivial winding of the phase of the order
parameter $\psi$.

The number density of the gas is $\hat{n} = \psi^\dagger \psi$ and can be written as a
sum of the average density $\nave$ and a fluctuation $\dn$: $\hat{n}(x,t) = \nave +
\dn(x,t)$. We consider the uniform case, when $\nave$ (the average) 
is independent of $x$, but
will comment briefly on the effect of nonuniform $\nave$, such as resulting from
a confining potential.

At large wavelengths, fluctuations of the density are small, so 
$\psi$ has well-defined modulus and phase. We write 
$\psi = (n+\dn)^{1/2} \exp(i\theta)$ and, expanding in small $\dn$, obtain
\be
L_E \approx i(\nave + \delta n) \partial_\tau \theta 
+ {n \over 2 m} (\partial_x \theta)^2 + {g \over 2} (\delta n)^2 \; .
\label{L1.5}
\ee
Note that we impose no restrictions on the size of fluctuations of $\theta$,
in accordance with the absence of long-range order.
Note also that in (\ref{L1.5}) we have omitted a term containing gradients of 
$\delta n$. This is correct in the leading longwave approximation, but at shorter 
wavelengths
that term will convert the purely acoustic dispersion law into the full dispersion
of Bogoliubov's quasiparticles \cite{Bogoliubov}.

The approximation (\ref{L1.5})
applies away from the instanton core, but not at the core itself.
This is because the size of the core is determined by the healing length
\be
\xi=(4gm \nave)^{-1/2} \; ,
\label{xi}
\ee
while the long-wavelength approximation corresponds to wavenumbers $k \ll \xi^{-1}$
and so does not resolve the core.

In this approximation,
the density fluctuations are Gaussian, and can be integrated out in the path
integral. This amounts to using the equation of motion for $\delta n$
and substituting the result, $\delta n = -(i/g) \partial_\tau \theta$, back into
eq. (\ref{L1.5}). In this way, we obtain a phase-only theory with the 
Euclidean Lagrangian
\be
L_E \approx i\nave \partial_\tau \theta + {1\over 2g} (\partial_\tau \theta)^2
+ {n \over 2 m} (\partial_x \theta)^2 \; .
\label{L2}
\ee
Instantons are singular solutions of this theory. The presence of such singular 
solutions means that, even though the Lagrangian (\ref{L2}) is quadratic in $\theta$, 
the theory remains non-Gaussian.

From eq. (\ref{L2}), we can read off the speed of Bogoliubov's phonons. 
We use units in which this speed is equal to 1:
\be
c_s = (gn /m)^{1/2} = 1 \; .
\label{vs}
\ee
In these units, eq. (\ref{xi}) takes the form
\be
\xi = (2gn)^{-1} \; .
\label{xi1}
\ee

If it were not for the first term, the Lagrangian (\ref{L2})
would be that of the usual XY model. In that theory, individual QPS can occur 
at $T=0$ in
any current-carrying state, due to the energy released by unwinding 
the supercurrent. This applies regardless of the strength of the coupling $g$.
Moreover, at a sufficiently strong coupling (which is outside the domain
of our semiclassical method) instanton-antiinstanton (IA)
pairs unbind even in vacuum, resulting in a BKT phase transition. 

However, the first term
in (\ref{L2}) drastically modifies the properties of the theory. We refer to this
term as {\em topological}. It is of the same nature as the topological contribution
\cite{Magnus}
to the Magnus force acting on vortices in higher dimensionalities (2D and 3D).
As we will now see, in 1D, for Euclidean paths that contain 
instantons, the topological term gives a nonzero contribution to the Euclidean action.
Being purely imaginary, that contribution can be interpreted as
the interference phase between QPS at different locations. 

The simplest case when the the topological term can be seen to play a role
is a single IA pair, with instanton located at $x^\mu_0=(x_0,\tau_0)$ and
antiinstanton at ${x^\mu_0}'=(x'_0,\tau'_0)$. 
Away from the cores, the density is approximately $\nave$, and the phase,
in the absence of supercurrent, is
\be
\theta_{\rm pair}(x,\tau) = \arg[x - x_0 +i (\tau - \tau_0) ]  
- (x^\mu_0 \leftrightarrow {x^\mu_0}') \; .
\label{pair}
\ee
For periodic boundary conditions, this is an approximation, which applies when
all of the distances involved: $|x^\mu - x^\mu_0|$, $|x^\mu - {x^\mu_0}'|$, 
and $|x^\mu_0 -{x^\mu_0}'|$, 
are much smaller than the length $L$ of the system.

The time derivative of the configuration (\ref{pair}) is
\be
\partial_\tau \theta_{\rm pair} = \frac{x-x_0}{(x-x_0)^2 + (\tau-\tau_0)^2}
- (x^\mu_0 \leftrightarrow {x^\mu_0}') \; .
\ee
Integrating over $\tau$ and $x$, we obtain the topological action:
\be
\Delta S_{\rm pair} = - 2\pi i \nave (x_0 - x'_0) \; .
\label{DS}
\ee
Since the variable conjugate to the position is momentum, we can interpret the action
(\ref{DS}) as resulting from production of momentum 
\be
P = 2\pi \nave \; ,
\label{P}
\ee
by the antiinstanton and its absorption by the instanton.

The above interpretation is of course nothing new and is 
readily confirmed by a direct calculation. Noether's
momentum following from the real-time version of eq. (\ref{L1}) is
\be
P = -i \int dx \psi^\dagger \partial_x \psi \; .
\ee
The difference between the initial and final momenta can be written as an integral
over a large closed contour in the $(x,\tau)$ plane:
\be
P_{\rm fin} - P_{\rm ini} = -i \oint_C d{\bf l} \cdot  \psi^\dagger \nabla \psi \; ,
\label{Pdiff}
\ee
where $C$ runs clockwise. If the contour encloses a single instanton, 
without any phonon excitations, then away from the core the density is $\nave$, and
(\ref{Pdiff}) equals precisely $-2\pi \nave$.

\section{QPS rate at $T=0$}
\label{sect:T=0}
In general, the metastable states connected by the instanton are current-carrying,
which implies that the phase $\theta$ winds an integer number $N$ times over
the length $L$ of the system. A suitable generalization of (\ref{pair}) is then
\be
\theta_c(x, \tau) = \theta_{\rm pair}(x, \tau) + \frac{2\pi}{L} N x \; .
\ee
The second, winding, term corresponds to superfluid velocity
\be
V = \frac{2\pi N}{m L} \; .
\label{svel}
\ee
In what follows, we will always assume that $V$ is much smaller than the speed
of sound $c_s = 1$ (although the more general case can be considered by similar
methods). For $V \ll 1$, the field of the IA pair is only weakly affected by the
superflow, and we can continue to use expression (\ref{pair}). However, the presence
superflow leads to an additional contribution to the action.
To logarithmic accuracy, we obtain
\be
S_{\rm pair} = -i P \dx + E  \dtau
+ \frac{2\pi}{g} \ln\frac{d}{\xi} + O\left( \frac{1}{g} \right) \; ,
\label{Spair}
\ee
where $\dx = x_0 - x'_0$, $\dtau = \tau_0 - \tau'_0$,
$d = (\dx^2 + \dtau^2)^{1/2}$ is the instanton-antiinstanton
separation, $\xi$ is the healing length (\ref{xi}), and
\be
E = E_N \equiv \frac{(2\pi)^2 N \nave}{m L} = P V \; .
\label{E}
\ee
Note that $E_N$ is precisely the energy released by a single instanton, as it 
unwinds the order parameter and reduces the supercurrent.

The use of logarithmic accuracy means that we assume the separation $d$
to be much larger than the instanton core size: $d \gg \xi$. This condition has
to be verified a posteriori for typical configurations. 

To compute the rate of QPS at zero temperature,
we need to integrate over all values of $\dx$ and $\dtau$:
\be
\rate \sim {\rm Im} \int d\dtau d\dx \exp(-S_{\rm pair}) \; .
\label{T=0}
\ee
The rate computed in this way will be the inclusive rate, i.e., it will take
into account the possibility of quasiparticle production in the final state.
In other words, eq. (\ref{T=0}) can be viewed as an optical theorem for
inelastic tunneling. This relation between the imaginary part of the partition 
sum of IA pair and the rate of inelastic tunneling \cite{Zakharov&al}
is familiar from the theory
of instanton-induced cross-sections in particle physics \cite{instantons}.
Another way to look at it is to think about the IA pair as a {\em bounce} 
\cite{Coleman},
describing the decay of a metastable current-carrying state. 
Then, eq. (\ref{T=0}) is the usual expression for the decay rate \cite{CC}, 
adapted to take into account the presence of the ``soft'' collective coordinates
$\dtau$ and $\dx$ corresponding to the IA separation.

As it is written, eq. (\ref{T=0}) does not include the preexponent and can be 
used to calculate only the leading exponential factor in the rate. 
(We will describe how to estimate the preexponent later.)
Assigning the main role to the exponential factor implies the use of
a semiclassical approximation and requires that the coupling $g$
should be small. However, we will see that
the main result---the absence of QPS in a uniform system at 
$T=0$---can be plausibly extrapolated to larger couplings.

The integral over $\dx$ is standard \cite{tables}:
\be
\int_0^\infty \frac{\cos(Px) d \x}{(\x^2 + \dtau^2)^{\nu + \half}}  = 
\left(\frac{P}{2|\dtau|}\right)^\nu 
\frac{\sqrt{\pi}}{\Gamma(\nu + \half)} K_\nu(P|\dtau|) \; ,
\label{integ}
\ee
where for our case $2\nu + 1 = 2\pi / g$.
Eq. (\ref{integ}) has to be integrated further over $\dtau$ with 
$\exp(-E\dtau)$. We see that this integral
is convergent for $E < P$ and divergent for $E > P$. This means that at
$E < P$ the IA pair has a finite separation, that is, QPS are always
bound in pairs and cannot occur individually. Note that the potential binding
QPS at $E< P$ is linear in $|\dtau|$.

Interpretation of the threshold at $E=P$ is simple. 
Since instantons produce momentum,
in a uniform system an isolated instanton has to be accompanied by production
of quasiparticles (Bogoliubov's phonons) that carry that momentum away, so that
momentum conservation is satisfied. But producing phonons with total momentum
$P$ requires, in a weakly-coupled gas, energy of at least $E=P$. Note that
the condition of weak coupling is essential here. The coherent state of phonons
that forms as a result of tunneling decays into individual quasiparticles, 
and in the weakly-coupled case we know their dispersion law---it is given
by Bogoliubov's formula \cite{Bogoliubov}. So, we can explicitly verify that
the energy of such a final state always exceeds its momentum (times $c_s$). 
This in general will not be true in a strongly interacting system (e.g., liquid
helium).

Now, if $E$ is given by eq. (\ref{E}), i.e., the only source of energy is unwinding
of the supercurrent, the condition $E > P$ can never be satisfied in the superfluid
state. This is obvious in the limit $V \ll 1$, in which we have derived the pair
action (\ref{Spair}), but the expression (\ref{E}) in fact holds also at larger $V$.
We see that $E > P$ implies $V > 1$, while according to Landau's criterion the 
superfluid state must have $V < 1$. Thus, in the absence of additional
sources of energy, the superfluid state is always in the regime 
$E < P$, and at $T=0$ individual QPS cannot occur. 

Although we have obtained this result within the weak-coupling limit,
the simple energetics underlying it allows us to speculate that it extends to
arbitrary values of $g$. In other words, unlike the XY model, 
the uniform 1D Bose gas has no BKT phase transition. This conclusion matches 
the observation that the correlation
function of the Bose gas, found in ref. \cite{Haldane}, evolve continuously 
with $g$. It is also consistent with the expected Galilean invariance of the 
$T=0$ state.

The requirement for vortex unbinding, $E > P$, 
can be avoided in a nonuniform system, where momentum
conservation need not be exact, for example, as a result of a short-scale
perturbation. For a weak perturbation with length scale of order of or shorter 
than $\xi$, the exponential factor in the rate, to the same logarithmic accuracy,
can be found by inserting a delta-function of $\dx$ under the integral in 
(\ref{T=0}), so that
\be
\rate_{\rm loc} \sim {\rm Im} 
\int d\dtau \exp\left( - E\dtau - \frac{2\pi}{g} \ln \frac{\dtau}{\xi} \right) \; .
\label{rate_local}
\ee

Eq. (\ref{rate_local}) coincides with the instanton rate that has been computed,
for a microcanonical initial state of energy $E$, in ref. \cite{periodic}. 
That computation has been 
done in the context of the Abelian Higgs model, in the limit when it effectively 
reduces to the XY model. The salient point of the calculation is that the
integral (\ref{rate_local}) is divergent 
(which, as we have seen, corresponds to unbinding of
the IA pair) and has to be defined by analytical continuation. The analytical
continuation produces a finite imaginary part. (For a similar discussion in
the context of dissipative quantum mechanics of a single degree of freedom, see
Ref. \cite{WG}.)

\begin{figure}
\leavevmode\epsfxsize=3.25in \epsfbox{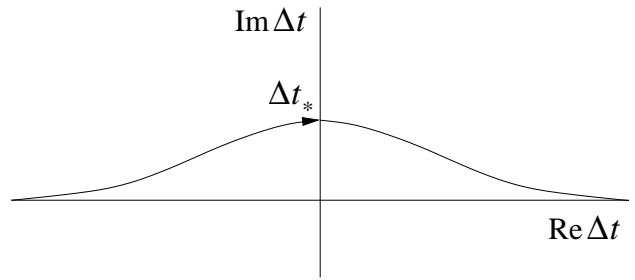}
\vspace*{0.2in}
\caption{
Integration contour passing through the saddle point (\ref{sp_local}).
}
\label{fig:tcont}
\end{figure}

Thus, we formally 
continue (\ref{rate_local}) to real-time separation $\dt = -i \dtau$:
\be
\rate_{\rm loc} 
\sim \int d\dt \exp\left(  -i E\dt - \frac{\pi}{g} \ln \frac{-\dt^2}{\xi^2}
\right) \; ,
\ee
and then observe that the integrand has a saddle point at
\be
\dt = \dt_* \equiv i\frac{2\pi}{gE} \; .
\label{sp_local}
\ee
Deforming the integration contour so that it passes through the saddle point,
as shown in Fig. \ref{fig:tcont}, and replacing the exponent with the saddle-point
value, produces the following exponential factor \cite{periodic}:
\be
\rate_{\rm loc}
\sim \exp\left\{ \frac{2\pi}{g} \ln (gE \xi) \right\} \; .
\label{rl1}
\ee
The exponent here has logarithmic accuracy, meaning that (\ref{rl1}) applies only
as long as $gE\xi \ll 1$.

Although this section is devoted primarily to the $T=0$ case, 
we also list here for future reference the counterpart of eq. (\ref{rl1}) for
$T \neq 0$. This can be obtained from the action of the periodic instanton of
Ref. \cite{periodic} by replacing the period with $\beta = 1/T$, or by 
integrating (\ref{rl1}) over $E$ with the Boltzmann factor $\exp(-\beta E)$.
The result reads
\be
\rate_{\rm loc} \sim \exp\left\{ \frac{2\pi}{g} \ln (T \xi) \right\}
\label{rl2}
\ee
and applies at $T\xi \ll 1$. In the context of cold Bose gases, perturbed by an
external potential at a length scale shorter than $\xi$, this result
was obtained by a different method in ref. \cite{Kagan&al}, where in addition
the preexponent was estimated.

Because sharply localized perturbations that lead to these relatively large rates
are unlikely to occur naturally in trapped atomic gases, it makes sense to
inquire about gentler sources of momentum non-conservation, such as a smooth 
variation of the trap potential. Since this question is somewhat outside
the main subject of this paper, we will only address it qualitatively. Namely,
we return to the action (\ref{Spair}), but now we do not assume that $E$ and $P$
in it are given by the specific expressions (\ref{E}) and (\ref{P}). This 
describes qualitatively a situation when some of the momentum is absorbed by the
potential. Note that for a smooth potential typical values of $P$ will
be close, even though not exactly equal, to the value (\ref{P}). Nevertheless, the
regime $E >P$ can now be realized.

The same setup can be used to model the presence of an additional
source of energy: we simply increase $E$ relative to (\ref{E}). One possible such
source is time-dependence of the parameters, which, as discussed in 
the introduction,
is of interest for analog models of gravity. For example, decreasing the coupling
$g$ reduces the interaction energy, and one could imagine that some of the 
released energy becomes available for enhancement of QPS. Of course, there is no 
reason to think that the expression (\ref{Spair}) with $E>P$ will be literally 
applicable in this case, but one may hope that it will at least mimic some of 
the main features of the situation. 

As before, the integral over $\dtau$ is defined by analytical continuation to
real $\dt = -i \dtau$, so that eq. (\ref{T=0}) becomes
\be
\rate \sim \int  d\dt d\dx e^{iP\dx -i E \dt}
\exp\left(  - \frac{\pi}{g} \ln\frac{\dx^2 - \dt^2}{\xi^2} \right) 
\; .
\label{rate0}
\ee
For $E > P$, the integrand has a saddle point at
\be
(\dt, \dx) = (\dt_*, \dx_*) \equiv i \frac{2\pi}{g} \frac{(E, P)}{E^2 - P^2} \; .
\label{sp}
\ee
Deforming the integration contours so that they pass through the saddle point,
cf. Fig. \ref{fig:tcont}, we obtain the exponential factor in the rate as
\be
\rate \sim \exp\left\{ \frac{\pi}{g} \ln [(E^2 - P^2) g^2 \xi^2 ] \right\} \; ,
\label{rateE}
\ee
where the exponent again has logarithmic accuracy.

We reiterate that, just as all the other rates computed in this section,
eq. (\ref{rateE}) is an inelastic rate---it corresponds to production of phonons
with total momentum $P$ in the final state. It can be viewed as a generalization
of eq. (\ref{rl1}) to the case when there is a nonzero 
transfer of momentum to
quasiparticles. Eq. (\ref{rateE}) has the expected threshold at $E=P$,
reflecting the requirement that production of phonons with momentum $P$ takes 
at least $E=P$ of energy (in units where the speed of phonons in equal to 1). 
At $E < P$, the integral in (\ref{T=0}) is convergent, has no imaginary part, and
the rate is zero. Note that the threshold is exponential: if the additional sources
of energy and momentum are weak, the difference $E - P$ is small, and the rate
is strongly suppressed.

\section{QPS rate at $T\neq 0$}
\label{sect:Tneq0}
\subsection{A summary}
We now turn to the main subject of this paper: calculation of the QPS rate 
in a uniform system at a low, but
finite, temperature $T$. The boundary problem, satisfied by the $S$-matrix
in the one-instanton sector at $T\neq 0$, is quite involved, and so we 
begin with a brief summary of the main points of the calculation.

We consider only the case of low temperatures,
\be
T \ll g n = \frac{1}{2\xi} \; .
\label{lowT}
\ee
It would be interesting to extend the calculation to higher $T$. In the
case of a sharply localized perturbation, we expect a crossover to thermal
activation at $T\sim gn$, i.e., when the period $\beta = 1/T$ of the periodic 
instanton becomes comparable to the core size $\xi$, cf. ref. \cite{periodic}.
In the Abelian Higgs model, such a crossover was indeed found numerically  
\cite{Matveev}.
In a uniform Bose gas, where momentum transfer 
is necessary, it is not clear how such a crossover will occur and, in fact,
even if it exists at all.
In this case, the region $T \sim gn$ can presumably be also addressed numerically,
but in the present paper we will limit ourselves to a few
comments on it in the conclusion.

An intuitive approach to the problem at temperatures (\ref{lowT}) would be
to place the system on a cylinder of circumference $\beta = 1/T$, i.e., consider 
configurations
that are periodic in the Euclidean time $\tau$ with period equal to $\beta$.
We have taken this approach in ref. \cite{slip}. However, to make this approach 
rigorous, one has to prove that the configurations found in 
ref. \cite{slip} are indeed the dominant pathways for phase slips.
A more systematic approach, which we take in this paper, 
is to first consider tunneling from a microcanonical
state of a given energy $E$ and then integrate over $E$ with the Boltzmann weight
$\exp(-\beta E)$. This second method allows us to find directly the energies 
corresponding to the dominant paths, and in particular to show that in the regime
(\ref{lowT}) thermally-assisted tunneling is the dominant phase-slip mechanism, 
more important than over-barrier activation (if any).

In addition, and curiously so, it turns out that to 
properly explore the initial and final states connected by tunneling, and also
to compute the first correction to the semiclassical exponent, requires using
a more precise dispersion law for quasiparticles than the simple acoustic one.
The method we use below allows us to take this modification into account
without having to explicitly find the instanton solution corresponding to the more 
precise dispersion law. This method, developed in ref. \cite{periodic}, is
a sort of perturbation theory in energy---hence the limitation to the low-$T$ 
range (\ref{lowT}), where the characteristic energies are not too large.
In the present paper, we adapt this method to the case when instantons transfer
momentum to quasiparticles.

Now, the main difference from tunneling at $T=0$ is that at $T\neq 0$ there
are preexisting quasiparticles in the initial state, and the tunneling path
can make use of those. Indeed, we will show explicitly that
tunneling now occurs between coherent quasiparticle states with total momenta
 $-P/2$ and $P/2$, so that the full momentum $P$ is transferred but only
energy $E\approx P/2$ is required. This is in contrast to the $T=0$ case, where 
quasiparticles with total momentum $P$ had to be produced, requiring energy of
at least $E=P$. Accordingly, the leading exponential in the rate at $T\neq 0$ is
\be
\rate_T \sim 
\exp(-P/2T + {\rm corrections}) \; ,
\ee
where the corrections are controlled by the small parameter $T/gn$. Using
the improved dispersion law for quasiparticles will allow us to obtain
the first of these corrections.

\subsection{The boundary problem and the modified instanton}

The starting point point of our calculation is
the expression \cite{periodic} for the microcanonical density matrix of phonons
\be
{\cal P}_E \sim \delta(\hat{H} - E)
\label{PE_def}
\ee
in the coherent-state representation:
\be
{\cal P}_E(b^*, a) = {\cal N} \int d\eta 
\exp\left( -iE\eta + \int dk b^*_k a_k e^{i\omega_k \eta} \right) \; .
\label{PE}
\ee
The constant normalization factor ${\cal N}$ is chosen so that ${\cal P}_E$ has
the property of a projector:
\be
{\cal P}^2_E(b^*, a) = {\cal P}_E(b^*, a) \; ,
\label{proj}
\ee
This makes it a projector on the subspace of a given energy $E$.

\begin{figure}
\vspace*{0.2in}
\leavevmode\epsfxsize=3.25in \epsfbox{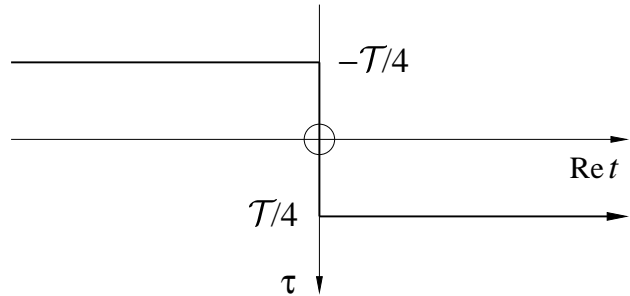}
\caption{
Complex time contour for calculation of the $S$-matrix in the one-instanton
sector. The circle denotes the instanton position.
}
\label{fig:cont}
\end{figure}

The density matrix (\ref{PE}) will be our initial condition 
on a complex time contour shown in Fig. \ref{fig:cont}, cf. ref. \cite{periodic}. 
The precise form of the contour will only become important later. For now,
all that matters is that there is some initial Euclidean time $\tau_i$ and
some final $\tau_f$, both of which can be complex but can be
associated with the distant past and distant future, respectively. 
As typical in problems where the evolution starts 
in the distant past, the interaction is adiabatically switched off in the
initial state, so that the Hamiltonian $\hat{H}$ in (\ref{PE_def}) is simply the 
Hamiltonian of free phonons. Accordingly, $\exp(i\omega_k \eta)$ factor in (\ref{PE}) 
is a result of the free evolution of a coherent state $|a\rangle$:
\be
e^{i \hat{H} \eta} |a_k \rangle = |a_k e^{i\omega_k \eta} \rangle \; ,
\label{free_evol}
\ee
where $\omega_k$ is the frequency of the phonon mode with momentum $k$.
The coherent states are normalized by the condition
$\langle b | a \rangle = \exp(b^* a)$. Combining this condition with 
eq. (\ref{free_evol}), one can in fact quickly derive eq. (\ref{PE}).

Note that we fix by hand the initial energy, but not the initial momentum. 
The most probable momentum of the initial state will be determined dynamically:
it is the total momentum of the state $\prod_k |a_k\rangle$ corresponding to 
the most advantageous tunneling path.

The probability of an individual QPS is
\be
{\rm Prob} = \Tr [S {\cal P}_E S^\dagger] = \Tr [S_E S_E^\dagger] \; ,
\label{prob}
\ee
where $S$ is the $S$-matrix, 
and $S_E = S{\cal P}_E$; we have used the projector property (\ref{proj}).
Tracing in eq. (\ref{prob}) is over all states that differ by unit of winding 
number from the supercurrent state on which the density matrix (\ref{PE}) is 
built. In other words, (\ref{prob}) is an inclusive probability.

In the coherent-state representation, $S_E$ acquires a convenient form 
\cite{periodic}:
\begin{widetext}
\be
S_E(b^*, a) = \int d\theta_i d\theta_f {\cal D} \theta d\eta
\exp\left\{ -i E\eta + B_i(a_k e^{i\omega_k \eta}, \theta_i) 
+ B_f(b_k^*, \theta_f) - \int_{\tau_i}^{\tau_f} L_E d\tau \right\} \; .
\label{SE}
\ee
\end{widetext}
Note that besides the usual path integral over the intermediate values
of the field $\theta$ and the integral over $\eta$, inherited from the
projector (\ref{PE}), we have here also integrals over the initial and final values
of the field. These are needed to convert $S_E$ to the coherent-state 
representation: $\exp{B_i}$ and $\exp{B_f}$ are the wave functions for the coherent
states $|a_k e^{i\omega_k \eta}\rangle$ and $\langle b|$:
\begin{widetext}
\ba
B_i(a_k e^{i\omega_k \eta}, \theta_i) & = &  
\int dk \left\{ -\half a_k a_{-k} e^{2\ok (-\tau_i + i\eta)} 
-\half \ok \ttheta_i(k) \ttheta_i(-k) 
+ \sqrt{2\ok} a_k \ttheta_i(k) e^{\ok (-\tau_i + i\eta)} \right\} \; , 
\label{Bi} \\
B_f(b_k^*, \theta_f) & = &
\int dk \left\{ -\half b^*_k b^*_{-k} e^{2\ok \tau_f} 
-\half \ok \ttheta_f(k) \ttheta_f(-k) 
+ \sqrt{2\ok} b^*_k \ttheta_f(-k) e^{\ok \tau_f} \right\} \; ,
\label{Bf}
\ea
\end{widetext}
where tildes denote spatial (in our case, one-dimensional) Fourier transforms, for
example,
\be
\ttheta_i(k) =\ttheta(k, \tau_i) = 
\int \frac{dx}{\sqrt{2\pi}} \theta(x, \tau_i) e^{ikx} \; .
\ee

The restriction that the winding number changes by one means that in (\ref{SE}) 
we consider paths of the form
\be
\theta(x,\tau) =  \theta_I(x-x_0,\tau-\tau_0) + \rg \nu(x,\tau) 
+ \frac{2\pi}{L} Nx \; ,
\label{theta}
\ee
where $\theta_I$ is the one-instanton solution,
and $\nu$ is a fluctuation that has zero total winding (but may still include
IA pairs). As before, we consider only relatively small $N$, when
the superfluid velocity (\ref{svel}) satisfies the condition $V \ll 1$.

The approximate instanton solution, obtained from the longwave Lagrangian 
(\ref{L2}), is
\be
\theta_I(x-x_0, \tau-\tau_0) = \arg[x-x_0 +i(\tau-\tau_0)] \; .
\label{inst}
\ee
We will see that at temperatures in the range (\ref{lowT}), the typical 
wavelengths
of quasiparticles participating in the process are indeed large, much larger than
the healing length $\xi$, so the longwave limit (\ref{L2}) is applicable. 
Nevertheless, there are corrections to results obtained in this limit.
In particular, for an
accurate study of the saddle point that determines the QPS rate, the instanton
(\ref{inst}) will have to be corrected (``modified''), to take into account 
a deviation of the quasiparticle dispersion law from the purely acoustic one.

Using (\ref{L2}) for $L_E$ in (\ref{SE}), we find
that the integration over $\nu(x,\tau)$ is Gaussian and gives 
rise to the free equation of motion:
\be
\nabla^2 \nu = 0 \; .
\label{eqnu}
\ee
The integrals over the boundary values of $\nu$ result in the boundary 
conditions (b.c.)
\ba
\omega_k \tnu_i(-k) - \dot{\tnu}_i(-k) & = & 
\sqrt{2\omega_k} e^{\omega_k (-\tau_i + i \eta)} a_k \; , \label{bc1}\\ 
\omega_k \tnu_f(k) + \dot{\tnu}_f(k) & = &
\sqrt{2\omega_k} e^{\omega_k \tau_f} b^*_k \; . \label{bc2}
\ea
Dots denote derivatives with respect to the Euclidean time $\tau$.
Thus, the coherent-state parameters $a$ and $b^*$ determine, through the b.c.
(\ref{bc1}) and (\ref{bc2}), the non-Feynman parts
of the fluctuation $\nu$---the negative frequency part at $\tau_i$ and the
positive frequency part at $\tau_f$.

The Fourier transform of the instanton field (\ref{inst}) itself
is computed at real $\tau_0$, such that 
${\rm Re} \tau_i < \tau_0 < {\rm Re} \tau_f$, and real $x_0$ 
and then analytically continued to arbitrary complex values. We find
\ba
{1\over \rg} \widetilde{\theta}_I(k, \tau_i) & = & 
\frac{1}{\sqrt{2\ok}} e^{\ok(\tau_i -\tau_0) + ik x_0} R(k) \; , 
\label{theta1} \\
{1\over \rg} \widetilde{\theta}_I(-k, \tau_f) & = & 
\frac{1}{\sqrt{2\ok}} e^{-\ok(\tau_f -\tau_0) - ik x_0} R(k) \; , 
\label{theta2}
\ea
where 
\be
R(k) = \frac{i}{k} \left(\frac{\pi\ok}{g} \right)^{1/2} \; .
\label{R}
\ee
Because the instanton field satisfies Feynman b.c. in both directions, it 
does not directly participate in the b.c. (\ref{bc1}), (\ref{bc2}). However, 
as seen from (\ref{Bi}), (\ref{Bf}), it acts as a source
for the coherent-state parameters $a$ and $b^*$.

In the expression (\ref{R}),
\be
\omega_k = |k| \; ,
\label{acous}
\ee
in accordance with the fact that the solution (\ref{inst}) was obtained from the
longwave limit (\ref{L2}), in which the phonon dispersion is a simple acoustic one.
As it turns out, important phonon momenta in our case are of the order
\be
k \sim (\xi^2 \per)^{-1/3} \ln^{1/3} \frac{\per}{\xi} \; ,
\label{kest}
\ee
where $\per \gg \xi$ is the period of the configuration (see below). Since this
momentum is much smaller than $\xi^{-1}$, the acoustic approximation (\ref{acous})
is adequate for use in the coefficients $R(k)$. However, as will be shown below,
in the exponentials $\exp(\pm \omega_k \tau)$ we need to use a more precise
approximation to Bogoliubov's full dispersion law \cite{Bogoliubov}:
\be
\omega_k = (k^2 + k^4 \xi^2)^{1/2} \approx |k| + \half |k|^3 \xi^2 \; .
\label{k3}
\ee
This means that the instanton that we will be using is in fact
modified relative to the simple expression (\ref{inst}). This modified instanton
could in principle be obtained from the Fourier transforms (\ref{theta1}),
(\ref{theta2}), in which we substitute the more accurate dispersion law 
(\ref{k3}) in the exponents. One of the advantages of the present approach, however, 
is that it will allow us
to obtain many interesting physical quantities without ever needing the explicit
form of the modified instanton.

\subsection{Microcanonical rate}
The solution to the boundary problem (\ref{eqnu})--(\ref{bc2}) is
\ba
\nu(x,\tau) & = & \int \frac{dk}{2\sqrt{\pi\ok}} \left(
a_k  e^{\omega_k (-\tau + i \eta) + ikx} + b^*_k e^{\omega_k \tau - ikx} 
\right) \nonumber \\
& + & \nu'(x,\tau) \; ,
\label{nu}
\ea
where $\nu'$ is a solution to (\ref{eqnu}) with Feynman b.c. A nontrivial $\nu'$
is only possible because the region of applicability of (\ref{eqnu}) has a ``hole''
at the instanton core. To find $\nu'$, we need, in principle, to consider
the equation for the entire fluctuation of the order parameter, including both 
the modulus and the phase. In other words, instead of (\ref{theta}) we would write
\be
\psi = (\psi_I + \delta \psi) \exp(2\pi i N x / L) \; ,
\ee
where $\psi_I$ is the instanton solution, and $\delta \psi$ is a fluctuation.
Instead of the longwave limit (\ref{L2}), we would have to consider the full 
Lagrangian (\ref{L1}). 
The integration over $\delta \psi$ will no longer be automatically Gaussian, but
it will become such in the leading semiclassical approximation. The corresponding
equation for the fluctuation is
\be
{\cal D}^2 \delta \psi = 0 \; ,
\label{D2}
\ee
where ${\cal D}^2$ is the second-order differential operator 
in the instanton background. Fortunately, in what follows we will not need the
explicit form of (\ref{D2}), but only the general properties of its solutions.

There are two types of solutions to eq. (\ref{D2}). Solutions satisfying the
Feynman b.c. are the zero modes of the operator ${\cal D}^2$, associated with 
the collective coordinates $x_0$, $\tau_0$ of the instanton (\ref{inst}).
The coefficients with which these zero modes occur in $\delta \psi$ are arbitrary
and need to be integrated over. These integrations can be converted in the usual
way into integration over the collective coordinates. 

The other type of solutions are delocalized modes. These contain non-Feynman parts,
so the coefficients with which they appear in $\delta \psi$ are fixed by the b.c.
The main idea
of the perturbative method developed in ref. \cite{periodic} is that when the
typical momenta $k$ of excitations involved in tunneling are small, 
the delocalized modes can be approximated by the
plane-wave solution given by eq. (\ref{nu}) with $\nu' = 0$.
We stress that this ``perturbative'' method is not an expansion
in small coupling $g$. Instead, corrections to the plane wave, which are due
to scattering of the plane wave on the instanton core, are controlled by 
the parameter $k\xi$. According to the estimate (\ref{kest}), this parameter
is small. Thus, to the leading order, we can simply neglect $\nu'$ and use in 
(\ref{SE}) the field (\ref{theta}) with $\nu$ given by the plane wave only.

Next, we observe that the tracing in the expression (\ref{prob}) for the 
probability can be done in the coherent-state representation by integrating
over $a$ and $b$ with $\exp(-a^*a - b^* b)$ as the measure, and this integration 
is Gaussian. After some algebra, we obtain
\be
{\rm Prob} \sim \int dx_0 dt_0 dx'_0 dt'_0 d\eta d\eta' 
e^{-2 S_0 + i P \dx - i E_N \dt - i E\zeta +  W} ,
\label{prob2}
\ee
where
\be
W =
\int dk |R|^2 \Sigma(k) e^{i\ok\zeta} [e^{-i\k\X} + e^{i\k\X - i\ok \zeta} - 2] 
\; .
\label{W}
\ee
We have introduced the following notation:
$\zeta= \eta - \eta'$, $\dx = x_0 - x'_0$, $\dt = t_0 - t'_0$, 
$\k \X = \ok \dt - k \dx$, and
\be
\Sigma(k) = (1 - e^{i\ok\zeta})^{-1} \; .
\label{Sigma}
\ee
$R$ is given by eq. (\ref{R}). As in eq. (\ref{rate0}), we have regularized the
divergent integrals over $\tau_0$ and $\tau'_0$ by continuation to real time.

The doubling of the number of integrations in
(\ref{prob2}) has to do with the presence of two path integrals in (\ref{prob}):
one in $S_E$, and the other in $S_E^\dagger$. If $S_E$ is associated with an
instanton, then $S_E^\dagger$ can be associated with an antiinstanton. The first
three terms in the exponent of (\ref{prob2}) are the sum of the actions of a
single instanton and a single antiinstanton. In particular, to logarithmic
accuracy,
\be
S_0 = \frac{\pi}{g} \ln \frac{L}{\xi} \; ,
\label{S0}
\ee
where the size of the system $L$ is used as an
infrared cutoff. Comparing (\ref{S0}) to the action (\ref{Spair}) of
an instanton-antiinstanton (IA) pair, we see that $L$ now appears in place of the
IA separation $d$. This is because we are now computing not the action of a pair,
but the sum of the individual actions, each of which is infrared-divergent. 
As we will see, the dependence on $L$ will be removed by
the $W$ term (\ref{W}), whose presence reflects the non-vacuum nature of the
initial and final states of phonons. This is similar to how 
the IA interaction at $T=0$ reflects a non-vacuum final state, a
correspondence well-known from the studies of instanton-induced cross-sections
in particle theory \cite{Zakharov&al,periodic,instantons}.

Nontrivial integrals in (\ref{prob2}) are those over the relative positions
$\zeta$, $\dt$, and $\dx$. The remaining integrals, those over the 
``center-of-mass'' positions, simply produce powers of space and time volumes, 
which are either absorbed in the normalization of the projector ${\cal P}_E$, 
or factored out when we compute the probability per unit length and unit time,
i.e., the rate.

We begin by integrating over $\zeta$ and $\dt$. These integrals can be
done by steepest descent, and the corresponding saddle-point conditions have
simple physical meaning \cite{periodic}. In our case, the saddle-point 
conditions are
\ba
E & = & 2 \int dk \ok  |R|^2 e^{i\ok\zeta} (\cos\k\X - 1) \Sigma^2 \; ,
\label{condE} \\
E_N & = & \int dk \ok  |R|^2   
 (e^{iK\X} - e^{i\ok\zeta -iK\X} ) \Sigma \; . \label{condEN}
\ea
These can be rewritten as
\ba
E & = & \int dk \ok  a^*_k a_k \; , \label{condE2} \\
E_N & = & \int dk \ok  (b^*_k b_k - a^*_k a_k) \; , \label{condEN2}
\ea
where $a$, $a^*$, $b$, and $b^*$ are the saddle-point values corresponding
to the Gaussian integration that led to (\ref{prob2}). 
These values characterize the most
probable initial and final states and will be discussed in detail in the
next subsection. For now, we will only need the corresponding quasiparticle
densities:
\ba
a_k^* a_k & = & |R|^2 \frac{\sin^2[\ok \dt / 2 - k  \dx / 2]} {\sin^2(\ok \zeta /2)} 
\; , \label{dens1} \\
b_k^* b_k & = & |R|^2 \frac{\sin^2[\ok (\zeta-\dt) / 2 + k  \dx / 2]} 
{\sin^2(\ok \zeta /2)} \; . \label{dens2}
\ea
where the upper sign corresponds to $a$, and the lower one to $b$. Notice that
(\ref{condE2}) is a natural expression for the energy of the initial
state, while (\ref{condEN2}) expresses energy conservation: the change in energy
of the phonon subsystem equals the energy produced by unwinding the current.

It is of interest to consider both the case when $\Delta x$ is real (which is
its original domain), and the case when $\Delta x$ is purely imaginary (which is 
where the saddle-point for it will be found). In either of these cases, the
saddle points for $\zeta$ and $\dt$ are purely imaginary, in particular,
\be
\zeta = i \per \; ,
\label{zeta}
\ee
with $\per > 0$. As we will see, $\per$ is the period of the configuration.

In this paper, we consider only the limit when the energy $E_N$
released by unwinding the current is much smaller than the typical thermal energy
$E$ (although the more general case can be considered similarly). Then, the
left-hand side of (\ref{condEN}) can be set to zero, and we find that
\be
\dt = \frac{\zeta}{2} = \frac{i}{2} \per \; .
\label{t}
\ee
These are the same saddle-point relations as those found in ref. \cite{periodic}, but
obtained here for a somewhat more general situation---when an instanton causes
a nonzero momentum transfer.

The energy condition (\ref{condE}) can now be rewritten in the form
\be
E = \half \int_{-\infty}^{\infty}  \frac{dk \ok |R|^2}{\sinh^2 (\ok\per/2)}
[\cosh (\ok\per/2) \cos k\dx - 1] \; .
\label{ene_cond}
\ee
This implicitly determines $\per$ in terms of energy $E$.

The exponent $W$ at the saddle point equals
\be
W = \int_{-\infty}^{\infty} dk |R|^2 \left\{
\frac{\cos k\dx - 1}{\sinh(\ok \per / 2)}
+ \frac{2}{e^{\ok \per / 2} + 1} \right\} \; .
\label{Wsp}
\ee
The integral of the second term is infrared-divergent. 
Using the longitudinal size $L$ as
an infrared cutoff, we find that to logarithmic accuracy the integral is
$(2\pi / g) \ln(L/\per)$. In the exponent of eq. (\ref{prob2}) for the probability,
$W$ is combined with twice the instanton action
(\ref{S0}). As a result, the dependence on $L$ disappears. 

By a direct calculation, one can verify that the saddle-point expressions
(\ref{ene_cond}) and (\ref{Wsp}) are related:
\be
\left. \frac{\partial W}{\partial \per} \right|_{\dx} = -E \; .
\label{WandE}
\ee
This relation is convenient if we want to restore $W$ (up to a constant) 
from an already calculated $E$.

For real $\dx$, the first integral in (\ref{Wsp}) rapidly converges in the 
ultraviolet. If we use the simple acoustic dispersion law (\ref{acous}), 
this integral can be computed explicitly, and we obtain, to logarithmic accuracy,
\be
W_{\rm ac} = - \frac{2\pi}{g} \ln \cosh \frac{\pi \dx}{\per} 
+ \frac{2\pi}{g} \ln\frac{L}{\per} \; .
\label{Wac}
\ee
This is the same expression as obtained in ref. \cite{slip} by computing the 
action of a certain periodic field. 
For an individual real $\dx$, (\ref{Wac}) is indeed
an adequate approximation to $W$, but we still need to integrate over $\dx$. 
This will be done by steepest descent, and we will see that on the corresponding 
(complex) saddle point the integral does not converge as rapidly. As a result,
the saddle point cannot be thoroughly explored in the acoustic approximation: 
we will need the more accurate dispersion law (\ref{k3}).

Using eq. (\ref{Wsp}), we obtain the following saddle-point equation for $\dx$:
\be
iP = \int_{-\infty}^{\infty} dk k |R|^2 \frac{\sin k\dx}{\sinh(\ok \per / 2)} 
\; ,
\label{condP}
\ee
which shows that the saddle-point $\dx$ lies on the upper imaginary axis.
Eq. (\ref{condP})
can be cast into a form similar to eqs. (\ref{condE2}), (\ref{condEN2}):
\be
\frac{P}{2} = - \int_{-\infty}^{\infty} dk k a^*_k a_k 
= \int_{-\infty}^{\infty} dk k b^*_k b_k \; .
\label{condP2}
\ee
The integrals here are the total momenta of quasiparticles in the initial and 
final states. Thus, on the one hand, (\ref{condP2}) expresses momentum
conservation and, on the other, shows that tunneling occurs between states with
momenta $\mp P/2$.

We are
interested in the range of energies, for which the periods satisfy
\be
\per \gg \frac{1}{g n} = 2\xi 
\label{condT}
\ee
(the second relation applies due to our choice of units with $c_s = 1$).
In this case, the saddle-point $\dx$ is of the form
\be
\dx = \frac{i}{2} (\per + \delta)
\label{x}
\ee
with $0 < \delta \ll \per$. Using the corrected dispersion law (\ref{k3}) 
in the exponents and the acoustic $\omega_k = |k|$ in the coefficient $R$,
we take eq. (\ref{condP}) to the form
\be
\frac{P}{2} = \frac{\pi}{g} \int_0^{\infty} dk 
\exp(-{1\over 4} k^3 \xi^2 \per + \half k\delta ) \; .
\label{condP3}
\ee
We see that without the cubic term in the exponent the integrand would not have 
the correct large-$k$ behavior. 

The integrand in (\ref{condP3}) has a maximum at $k=k_*$,
\be
k_*^2 = \frac{2 \delta}{3\xi^2 \per} \; .
\label{k_*}
\ee
Assuming that the maximum is sufficiently
sharp (this can be confirmed a posteriori), and approximating the integrand 
near it with a Gaussian, we obtain
\be
k_*^3 = \frac{2}{3\xi^2 \per} \left\{
\ln\frac{\per}{\xi} + O(\ln\ln \frac{\per}{\xi}) \right\} \; .
\label{k_*3}
\ee
Here we have used $P= 2\pi n =\pi / g\xi$. The sharpness of the maximum 
and therefore the accuracy of this calculation is 
controlled by the large $\ln(\per /\xi)$.

Now, comparing the expressions (\ref{condE2}) for the energy and (\ref{condP2})
for the momentum, we see that the main difference between the two is
due to the deviation of $\omega_k$ from the strict acoustic form. Indeed, 
(assuming $E_N\to 0$) we can write
\be
E - {P\over 2} = \int_{-\infty}^{\infty} dk (\ok - k) b^*_k b_k \; .
\label{ok-k}
\ee
Using (\ref{dens2}) with the saddle-point values of $\zeta$, $\dt$, and $\dx$,
we see that for $k<0$ the integral is rapidly converging, and the contribution
from this region is small: most of the total comes from $k>0$, where the integral
converges much more slowly. Using the same Gaussian approximation as above, 
we obtain
\be
E - {P\over 2} = {1\over 4} \xi^2 k_*^2 P 
\left\{ 1 + O(\ln^{-1/2}\frac{\per}{\xi}) \right\} \; ,
\label{E-P/2}
\ee
or, substituting $k_*$ from (\ref{k_*3}),
\be
 E - {P\over 2} \approx 
\frac{P}{4} \left( \frac{2\xi}{3\per} \right)^{2/3} \ln^{2/3} \frac{\per}{\xi} 
\; .
\label{Eapprox}
\ee
Thus, the energy of the optimal initial state is close to $P/2$, but there
is a correction, given by the right-hand side of (\ref{E-P/2}).

Note that in the above calculation it was sufficient to use $R(k)$ obtained in the
limit $\ok = |k|$: any corrections to $R(k)$ due to the modified dispersion law
multiply the already small $\ok -k$ in (\ref{ok-k}) and do not affect the leading
correction computed in (\ref{E-P/2}). The same applies to any changes in the
relation between $a$, $b$ and $R$ that are due to scattering corrections.

We can now use (\ref{WandE}) to restore the exponent $W$ governing the QPS
rate. In doing so, we need to take into account the fact that in (\ref{WandE}) 
the derivative is at fixed $\dx$, while (\ref{E-P/2}) was obtained using 
the saddle-point $\dx$, which itself is a function of $\per$. This difficulty
can be circumvented in the following way. We first rewrite (\ref{WandE}) as
\be
\frac{\partial}{\partial \per} \left. 
\left( W + iP \dx \right) \right|_{\dx} = -E 
\label{WandE2}
\ee
and then observe that the partial derivative here can be replaced by the total,
since the part due to the dependence of $\dx$ on $\per$ vanishes at 
the saddle point.
Thus, integrating (\ref{E-P/2}) over $\per$ we obtain not just $W$ but 
the sum $W + iP\dx$. 

For the full exponent
\be
 \tW = -2S_0 + iP\dx - iE\zeta + W \; , 
\label{tWdef}
\ee
which according to 
eq. (\ref{prob2}) governs the exponential factor in the rate 
(in the limit $E_N \to 0$), we find
\be
\tW = (E-{P\over 2}) \per -
\frac{\pi}{2g} \left( \frac{3\per}{2\xi} \right)^{1/3} 
[1 + O(\ln^{-1/2}) ] \ln^{2/3} \frac{\per}{\xi} \; ,
\label{tW}
\ee
where $\ln\equiv \ln(\per/\xi)$.
In the same approximation, we can use (\ref{Eapprox}) to express the period $\per$
through energy:
\be
\per \approx \frac{\xi}{8} \left( \frac{P}{E- P/2} \right)^{3/2}
\ln \frac{P}{E- P/2} \; .
\label{per}
\ee
Substituting this into eq. (\ref{tW}), we obtain our final expression for the
exponent:
\be
\tW \approx -\frac{\pi}{4g}  \left( \frac{P}{E- P/2} \right)^{1/2}
\ln \frac{P}{E- P/2} \; ,
\label{tW2}
\ee
which applies at energies such that
\be
E - P/2 \ll P \; .
\label{condEfin}
\ee
The microcanonical rate is
\be
\rate \sim \exp (\tW) \; .
\label{rate_micro}
\ee
It corresponds to the optimal choice of the tunneling endpoint among all
states of a given energy $E$. It is exponentially larger that the estimate
(\ref{rateE}), which corresponds to some a priori, non-optimal, way of injecting
the energy. In particular, the threshold for quasiparticle production
has moved from $E=P$ in (\ref{rateE}) to $E= P/2$ in (\ref{rate_micro}).

\subsection{The initial and final states} \label{sect:ini&fin}
We have referred several times to $\per$ as the period, but we have not yet
exhibited the periodicity of the field configuration. The field can be
obtained by substituting the saddle-point values of $a$ and $b^*$ into 
eq. (\ref{nu}). In addition to determining the field, these parameters
also determine the most probable initial and final states. We now turn to a
detailed discussion of $a$ and $b^*$.

The requisite saddle-point values are
\ba
a_k & = & \frac{1-e^{\ok \dtau -i k\dx}}{1- e^{i\ok \zeta}} 
R^* e^{\ok \tau'_0 - i k x'_0 - i\ok \eta'} \; , \label{ak}  \\
b_k^* & = & \frac{e^{i\ok\zeta} - e^{\ok \dtau -i k\dx}}{1- e^{i\ok \zeta}}
R e^{-\ok \tau_0 + i k x_0} \label{bk} \; .
\ea
The values for $a^*$ and $b$ are obtained by reverting all signs in all
exponentials and replacing $R$ with $R^*$. 
Substituting these expressions into eq. (\ref{nu}) and setting $\nu' = 0$, 
we obtain the fluctuation $\nu$.

The saddle-point solution (\ref{t}) fixes the difference between $\tau_0$
and $\tau'_0$, but not the ``center-of-mass'' position $(\tau_0 + \tau'_0)/2$.
The latter will in general have both real and imaginary parts. The real part
is fixed by the position of the time contour, on which the $S$-matrix is defined.
If we use the contour shown in Fig. \ref{fig:cont}, then the real part of $\tau_0$
is zero, so we can write $\tau_0 = i t_R$, where $t_R$ is real---it is the moment
(in real time) at which the QPS tales place.
Since the collective mode associated with $t_R$ is not 
important for the present argument, we set $t_R=0$, so that
\ba
\tau_0 & = & 0 \; , \\
\tau'_0 & = & \per / 2 \; .
\ea
This places the instanton 
at the origin, as indicated in Fig. \ref{fig:cont}.

With the help of (\ref{ak}) and (\ref{bk}), the fluctuation $\nu$ is obtained
as an integral over $k$, see eq. (\ref{nu}).
To similarly represent the full field $\theta=\theta_I + \sqrt{g} \nu$, 
we need also the Fourier transform of the (modified) instanton
field $\theta_I$.
Note that this has different forms in the regions $\tau > 0$ and $\tau < 0$,
cf. eqs. (\ref{theta1}), (\ref{theta2}). For definiteness, we consider 
$\tau > 0$, where we can use eq. (\ref{theta2}) with $\tau_f$ replaced by 
$\tau$. As a result, the field can be written as
\ba
\lefteqn{
\frac{1}{\sqrt{g}} \theta(x,\tau) = 
} \nonumber \\ & &
\int \frac{dk}{2\sqrt{\pi\ok}} (e^{i\ok\zeta} - e^{\ok \dtau -i k\dx}) f(x,\tau)
\Sigma(k) \; ,
\label{theta_tot}
\ea
where
\be
f(x,\tau) =
R e^{\ok (\tau-\tau_0) - ik(x - x_0)} + R^* e^{-\ok (\tau-\tau'_0) + ik(x - x'_0)} \; ,
\label{f}
\ee
and $\Sigma(k)$ is defined in (\ref{Sigma}). Note that $\Sigma$ can be rewritten as
\be
\Sigma(k) = \sum_{n=0}^{\infty} e^{in\ok \zeta} \; .
\ee
This allows us to interpret (\ref{theta_tot}) as 
a sum over instantons and antiinstantons at various locations, in parallel with
the interpretation of periodic instantons in ref. \cite{periodic}.

Indeed, if we substitute $\zeta = i\per$ and restrict our attention to 
the interval $(0, \per / 2)$, we see that the field (\ref{theta_tot}) can be 
interpreted
as the sum of fields from two periodic chains: one of instantons, at locations
$x = x_0$, 
$\tau = n \per$, and the other of anti-instantons, at locations $x = x'_0$,
$\tau = (n + 1/2) \per$, $n=0,\pm 1, \ldots$.

If we keep $x_0$ and $x_0'$ real, we have a family of periodic 
configurations, such as the one shown in Fig. \ref{fig:inst}. 
These are the same configurations
as found in ref. \cite{slip} by imposing from the start the requirement of periodicity
in the Euclidean time $\tau$.
Here we have reconstructed them without any such a priori requirement, following 
instead the perturbative method of 
ref. \cite{periodic}. We have seen, however, that the integral over $\dx$ in the
probability (\ref{prob2}) is not determined by a real saddle point. So, unlike the case
considered in \cite{periodic}, none of these real-$x_0$ configurations is an approximate
classical solution. The approximate solution that determines the rate corresponds to
the complex saddle-point (\ref{x}) and is itself complex.

For real $x_0$, $x_0'$, the integral in (\ref{theta_tot}) 
converges in the ultraviolet, for any $\tau$ in the interval
$(0, \per /2)$, even in the acoustic approximation $\ok = |k|$. 
Using the acoustic dispersion is equivalent to approximating each instanton in 
the chain by $\theta_I(x - x_0, \tau - \tau_n)$, where $\theta_I$ is the unperturbed 
solution (\ref{inst}). The sum
over $n$ can then be done explicitly, resulting in \cite{slip}
\ba
\lefteqn{
\theta_{\rm ac}(x,\tau) 
= \frac{1}{2i} [ \ln(1-e^{-i\tau-x+x_0}) - \ln(1-e^{i\tau-x+x_0}) 
 } \nonumber \\
 & &   - \ln(1+e^{-i\tau-x+x'_0}) + \ln(1+e^{i\tau-x+x'_0}) ] \; ,
\label{theta_c}
\ea
where all distances and times are measured in units of $\per/2\pi$. This 
configuration is explicitly periodic with period $2\pi$, and its Euclidean
action per period
to logarithmic accuracy equals $-W_{\rm ac} - iP\dx + 2 S_0$, 
where $W_{\rm ac}$ is the approximate expression (\ref{Wac}).

Now, consider the case when 
\ba
x_0 & = & x_R + \dx / 2 \; , \label{x0} \\
x_0' & = & x_R - \dx / 2 \; , \label{x0'}
\ea
where $x_R$ is real, and $\dx$ is the saddle-point value (\ref{x}). Note that these
$x_0$ and $x_0'$ are complex conjugate to each other. In this case, the integral
in (\ref{theta_tot}) is not convergent in the acoustic approximation, and we need
to use the more precise dispersion law (\ref{k3}). 

The sections of Euclidean time at $\tau= \pm \per/4$, 
which are half-way between the instantons and
antiinstantons, are expected to have a special significance. Since the complete
periodic solution determines the rate (i.e., the amplitude squared), 
we should be able to cut it in half, at $\tau=  \pm \per/4$, to obtain 
the tunneling path, and then attach this tunneling path to real-time evolution.
A direct calculation shows that when $x_0$ and $x_0'$ are complex conjugate,
as in (\ref{x0}), (\ref{x0'}), the Euclidean velocity 
$\partial_\tau \theta$ at $\tau= \per/4$ is purely imaginary, 
while the field itself there is purely real.
The same is true at $\tau=- \per/4$. Therefore, the solution becomes purely real
at
\be
\tau = \pm \per / 4 + i t' 
\label{t_R}
\ee
with real $t'$, i.e., on the horizontal, real-time, segments of the contour of 
Fig. \ref{fig:cont}.

An immediate consequence of this result
is that the real-time segments do not contribute to the imaginary part of the action,
so the tunneling exponent is determined by the Euclidean segment alone.
Another consequence is that the real-time evolution can
be interpreted as formation
and decay of the coherent fields corresponding to the tunneling endpoints.
For example, at $\tau = \per / 4 + i t'$ the real-time solution, as a function
of $t'$ and $x$, can be read off the expression (\ref{theta_tot}), in which
we substitute eqs. (\ref{x0}), (\ref{x0'}) and the saddle-point values of all the
parameters. Note that due to the deviation of the dispersion law from the purely
acoustic one, the wave packets of quasiparticles do disperse, so at 
$t' \to \pm \infty$ the coherent states become collections of free quasiparticles.
This means that the expressions (\ref{condE2}) for the total energy and (\ref{condP2})
for the total momentum should apply quite generally, i.e., even with the scattering
corrections included, despite the fact that these corrections will modify the
simple relations (\ref{ak}), (\ref{bk}) between $a$, $b^*$ and the instanton's
Fourier transform.

Finally, we note that, as clear from the expressions (\ref{dens1}), (\ref{dens2})
for the quasiparticle densities, the initial state (at $t' \to -\infty$) contains
mostly phonons with positive momenta, while the final state (at $t' \to \infty$) 
mostly with negative.

\subsection{Canonical rate} \label{subsec:can}
The leading exponential factor in the QPS rate at a temperature $T=1/\beta$ 
is obtained
by integrating the microcanonical rate over energy $E$ with
the Boltzmann weight: 
\be
\rate_T \sim \int dE e^{-\beta E} \rate \sim \int dE e^{W_\beta} \; ,
\label{rate_beta}
\ee
where 
\be
W_\beta = \tW - \beta E \; .
\ee
Since all variational parameters in the expression (\ref{tWdef}) for $\tW$ 
are already at their saddle-point values, the total derivative of $\tW$
with respect to $E$ coincides with the partial, i.e.,
\be
\frac{d \tW}{d E} = \per \; .
\label{dtW/dE}
\ee
Thus, the exponential in (\ref{rate_beta}) quite generically has an extremum 
at the energy $E_*$ corresponding to $\per = \beta$.
This is consistent with the standard argument, according to which at a finite
temperature we should be looking for solutions that are periodic in $\tau$ with 
period $\beta$. However, in general, the extremum can be either a maximum or a 
minimum. In the present case, under the low-temperature condition $\beta \gg \xi$, 
the energy $E_*$ falls into the range (\ref{condEfin}), where we can apply 
eq. (\ref{tW2}). We find that 
\be
E_* \approx  \frac{P}{2}+ \frac{P}{4}  \left( \frac{2\xi}{3\beta} \right)^{2/3} 
\ln^{2/3} \frac{\beta}{\xi} \; .
\label{E_*}
\ee
is a maximum, and the only one in this range. Thus, levels with energies near
$E_*$ give the main contribution to thermally-assisted tunneling among all
levels in the range (\ref{condEfin}). This has to compete with transitions
(tunneling or over-barrier) from levels outside
the range (\ref{condEfin}), i.e., those for which $E-P/2 = O(P)$. However,
the transition rate for such levels is suppressed at least by the Boltzmann
factor $\exp(-E / T) = \exp(-c P/ T)$, where $c > 1/2$. On the other hand,
as we will soon see, tunneling from $E \approx E_*$ is suppressed, at 
$\beta \gg \xi$, mainly by $\exp(-P/2T)$, so at these temperatures it is more 
important.

All that remains, then, is to substitute $\per = \beta = 1/T$ 
into eq. (\ref{tW}). We obtain
\ba
\lefteqn{\rate_T \sim } \label{W_beta} \\ & &
\exp\left\{ - \frac{\beta P}{2} 
-\frac{\pi}{2g} \left( \frac{3\beta}{2\xi} \right)^{1/ 3} 
[1 + O(\ln^{-1/2}) ] \ln^{2/3} \frac{\beta}{\xi} \right\} \; , \nonumber 
\ea
which is our final result. Here $\ln = \ln(\beta/\xi)$; the preexponent is
estimated in the next subsection.

Recall that $P=2\pi n$, where $n$ is the average
density. So,
under the condition (\ref{lowT}), the first term in the exponent of (\ref{W_beta})
is much larger than the second. This term was obtained in ref. \cite{slip} from
the approximate expression (\ref{Wac}), which is based upon using the
acoustic dispersion law throughout. Obtaining the second, correction, term
requires, as we have seen, the use of the more precise dispersion law (\ref{k3}).
The second term becomes of the same order as the first at $\beta \sim \xi$, where
the present approximation breaks down.

\subsection{Estimate of the preexponent}
A naive dimensional estimate for the preexponent in the QPS rate 
would be $L/\xi^2$, where $L$ is the total length of the system. In actuality,
of course, there is a multiplicative correction to this estimate.
Note, however, that the estimate of accuracy in (\ref{W_beta}) implies that we 
have already
omitted a variety of subleading terms, such as, for instance,
\be
- 2 S_0 + \frac{2\pi}{g} \ln\frac{L}{\beta} 
\approx - \frac{2\pi}{g} \ln\frac{\beta}{\xi}
\ee
(which is the leading contribution in the case of a sharply localized perturbation,
but is only subleading in the uniform system). At small $g$,
this term, omitted in the {\em exponent} of the rate, is more important than
any correction to the naive preexponent that we may obtain. Nevertheless, we 
now present an estimate of the preexponent, given the traditional interest in 
values of the ``attempt frequency'' (which is what the preexponent represents).

The preexponent comes from the ratio of two determinants: one for small fluctuations 
near the periodic instanton, and the other for small fluctuations in vacuum---similarly
to the case of a bounce \cite{CC},
for a review, see ref. \cite{Vainstein&al}.
Near the periodic instanton, fluctuations can be expanded into normal modes, which
satisfy the eigenvalue equation
\be
{\cal D}_P^2 \psi_n = \lambda_n \psi_n \; ,
\label{modes}
\ee
where ${\cal D}_P^2$ is a small-fluctuation operator in the periodic-instanton
background. The modes $\psi_n$ should be periodic with the same period 
$\beta$. We can use a classification of modes similar 
to that used after eq. (\ref{D2}): there are modes that are localized in regions
of a spatial size of order $\xi$ around the individual instantons and
antiinstantons, and modes that are not. Delocalized modes with 
$\lambda_n \ll 1/\xi^2$ (i.e., with typical momenta much smaller than $1/\xi$),
as well as modes with $\lambda_n \agt 1/\xi^2$, both delocalized and 
localized (if any), contribute only a numerical factor of order unity to the
ratio of the determinants (cf. ref. \cite{Vainstein&al}), and we will not be
interested in this factor in what follows.

Large factors can only come from localized modes 
with $\lambda_n \ll 1/\xi^2$, corresponding to the ``soft'' collective
coordinates.
The periodic instanton has two strictly zero modes ($\lambda_n =0$), corresponding 
to translations of the entire configuration in space and time. These are the
translations described by the parameters $x_R$ and $t_R$ of subsect. \ref{sect:ini&fin}.
Each of these zero modes contributes
a normalization factor of order $1/\xi\sqrt{g}$, up to a power of $\ln(\beta / \xi)$,
times the total volume associated with the
corresponding collective coordinate. The temporal volume cancels out when we go 
from the probability to the rate, while the spatial volume $L$ remains, resulting
in the following zero-mode factor
\be
\Omega_1 \sim \frac{L}{g \xi^2} \; .
\label{Omega_1}
\ee
In addition, there are two quasi-zero modes, corresponding to changes in 
the {\em relative} position of the instanton and antiinstanton chains. 
These modes are described by the saddle-point parameters $\dx$ and $\dt$.
They have approximately the same normalization factors as the strictly zero modes, 
but their volumes are determined by the saddle-point integrations, i.e., by the
second derivatives of the exponent $W$, eq. (\ref{W}), 
with respect to $\dx$ and $\dt$.
These volumes are of order $(k_* P)^{-1/2}$ each (while the mixed second derivative 
vanishes), 
so the quasi-zero modes contribute the factor
\be
\Omega_2 \sim  (g \xi^2 k_* P)^{-1} \sim (\xi k_*)^{-1} \; ,
\label{Omega_2}
\ee
where $k_*$ is given by eq. (\ref{k_*3}) with $\per = \beta$. 
Multiplying $\Omega_1$ and $\Omega_2$, we obtain an
estimate for the preexponent:
\be
\Omega \sim \frac{L}{g \xi^2} \left( \frac{\beta}{\xi} \right)^{1/3} \; ,
\label{Omega}
\ee
which is accurate up to a power of $\ln(\beta / \xi)$.

\section{Conclusion}
In a uniform system, instantons that generate momentum by unwinding a persistent
current need to transfer a compensating momentum to quasiparticles. 
We have seen that these 
instantons have many interesting properties that are absent in cases when no
such momentum transfer is necessary. We have considered in detail the case of
a weakly-coupled 1D superfluid at temperatures $T \ll gn \sim c_s / \xi$, 
where $c_s$ is the phonon speed, and $\xi$ is the healing length. 
(In this section, we restore
$c_s$, which was set to 1 earlier in the paper.), 

On the theoretical side, perhaps the most
curious features of this case are that the instanton is complex, has no turning points, 
and yet its analytical continuation to the appropriate real-time segments
is real. On the real-time segments, the solution can be interpreted as formation 
and decay of coherent states of quasiparticles. These initial and final states have
opposite total momenta $\mp P/2$, allowing for a transfer of momentum $P$.

A possibility of experimental detection of QPS in narrow superfluid channels via
momentum imaging has been
mentioned in the introduction. Looking at our final
result (\ref{W_beta}), which to the leading order we can rewrite as
\be
\rate_T \sim \exp \left( -\frac{\pi c_s}{g} \frac{gn}{T} \right) \; ,
\label{rate_T}
\ee
and comparing it to its counterpart (\ref{rl2})
for the case of a sharply localized perturbation, we see that
the temperature dependence of the rate is much steeper in the uniform 
case: a logarithm of $gn / T$ in the exponent is now replaced by a power.
As the calculation makes clear, this additional suppression results from the need to
use states with relatively high energy $E\approx c_s P/2$.  
The ``dynamical'' suppression, controlled by the quantum overlap between the optimal
initial and final states, is in the uniform case only subleading.

It is of interest to extend the present results in several directions.
First, it would be interesting to extend them to higher temperatures,
$T \sim gn$. A striking property of the rate (\ref{rate_T}) is that, at any
$T \ll gn$, it is exponentially
{\em smaller} than the rate of thermal activation that one would obtain
within the LAMH theory \cite{LA,McCH}. Indeed, when the energy $E_N$ released
by unwinding the current is negligible
(the same limit, in which (\ref{rate_T}) was obtained), 
the LAMH rate is
\be
\rate_{\rm LAMH} \sim \exp \left( -\frac{4c_s n}{3 T} \right) \; .
\label{LAMH}
\ee
On the other hand, we have shown in subsect. \ref{subsec:can} that at $T\ll gn$
our instanton is the dominant path for phase slips, more important than any
thermal activation. The discrepancy
can be traced to the fact that the original LAMH theory makes no account
of momentum conservation. Recall that the LAMH saddle point, whose energy determines 
the rate (\ref{LAMH}), has order parameter that is nonvanishing
everywhere (except for the special case of exactly zero current). Usually,
one assumes that this saddle point is close to some time-dependent fluctuation, for
which the order parameter vanishes at some point, allowing for a phase slip.
Our results imply that in a uniform system, where there are no external ``sinks''
of momentum (such as impurities, etc.), this is not a good assumption, i.e.,
in this case the LAMH saddle point does not nucleate any phase-slip process. 
We have preliminary
numerical data supporting this conclusion. Accordingly, it is by no means clear
if, in a uniform system, there is a crossover to a thermally activated mechanism
for phase slips at {\em any} $T$.

Another direction in which one may be able to extend the present results is
systems of higher dimensionality (2D and 3D). There, the role of the topological 
term is taken over by the Magnus force, which suppresses tunneling in a rather
similar way \cite{Volovik}. In either case, the suppression can be seen as a result
of destructive interference between tunneling at different values of the
spatial coordinate. It is natural
to ask if in 2D and 3D this suppression can be circumvented by an inelastic mechanism
(production of phonons) similar to the one considered here.

One last case of interest we mention is that of BCS-paired
superfluids and superconductors. In that case, an additional channel of momentum 
production \cite{slip} becomes available. It is associated with zero modes of 
fermionic quasiparticles at the instanton core.
For superconductors, the problem is complicated by the scattering of quasiparticles
on disorder (and, in 1D and 2D, on the boundaries of the sample),
which alters the momentum balance. Nevertheless, this channel of inelastic tunneling
deserves a further study, especially in view of its possible relevance to experiments 
\cite{exp1,exp2} on superconducting nanowires.

This work was supported 
in part by the U.S. Department of Energy through Grant DE-FG02-91ER40681 
(Task B).


\begin{thebibliography}{99}
\bibitem{Haldane} F. D. M. Haldane, Phys. Rev. Lett. {\bf 47}, 1840 (1981).
\bibitem{Lieb&Liniger} E. H. Lieb and W. Liniger, Phys. Rev. {\bf 130}, 1605
(1963); E. H. Lieb, {\it ibid.}, p. 1616.
\bibitem{solitons} T. Tsuzuki, J. Low Temp. Phys. {\bf 4}, 441 (1971);
V. E. Zakharov and A. B. Shabat, Zh. Eksp. Teor. Fiz. {\bf 64}, 1627 (1973)
[Sov. Phys. JETP {\bf 37}, 823 (1973)].
\bibitem{slip} S. Khlebnikov, Phys. Rev. Lett. {\bf 93}, 090403 (2004)
[cond-mat/0311045].
\bibitem{Unruh} W. G. Unruh, Phys. Rev. Lett. {\bf 46}, 1351 (1981).
\bibitem{Garay&al}  L. J. Garay, J. R. Anglin, J. I. Cirac, and P. Zoller,
Phys. Rev. Lett. 85, 4643 (2000) [gr-qc/0002015].
\bibitem{Fedichev&al} P. O. Fedichev and U. R. Fischer, 
Phys. Rev. Lett. {\bf 91}, 240407 (2003) [cond-mat/0304342].
\bibitem{Barcelo&al} C. Barcel\'{o}, S. Liberati, and M. Visser,
Phys. Rev. A {\bf 68}, 053613 [cond-mat/0307491].
\bibitem{Little} W. A. Little, Phys. Rev. {\bf 156}, 396 (1967).
\bibitem{LA} J. Langer and V. Ambegaokar, Phys. Rev. {\bf 164}, 498 (1967).
\bibitem{McCH} D. McCumber and B. Halperin, Phys. Rev. B {\bf 1}, 1054 (1970).
\bibitem{TAPS} E. J. Mueller, P. M. Goldbart, and Y. Lyanda-Geller,
Phys. Rev. A {\bf 57}, R1505 (1998).
\bibitem{Kagan&al} Yu. Kagan, N. V. Prokofiev, and B. V. Svistunov,
Phys. Rev. A {\bf 61}, 045601 (2000).
\bibitem{Volovik} G. E. Volovik, Pis'ma ZhETF {\bf 65}, 201 (1997)
[JETP Lett. {\bf 65} 217 (1997)].
\bibitem{momen} Yu. G. Makhlin and G. E. Volovik, Pis'ma ZhETF {\bf 62}, 923
(1995) [JETP Lett. {\bf 62}, 941 (1995)].
\bibitem{exp1} N. Giordano, Phys. Rev. Lett. {\bf 61}, 2137 (1988); Physica B
{\bf 203}, 460 (1994).
\bibitem{exp2} A. Bezryadin, C. N. Lau, and M. Tinkham, 
Nature (London) {\bf 404}, 971 (2000); C. N. Lau {\em et al.}, 
Phys. Rev. Lett. {\bf 87}, 217003 (2001).
\bibitem{instantons} For reviews, see: 
M. P. Mattis, Phys. Rept. {\bf 214}, 159 (1992);
P. G. Tinyakov, Int. J. Mod. Phys. A {\bf 8}, 1823 (1993);
R. Guida, K. Konishi, and N. Magnoli, Int. J. Mod. Phys. A {\bf 9}, 795 (1994);
V. A. Rubakov and M. E. Shaposhnikov,
Usp. Fiz. Nauk {\bf 166}, 493 (1996) [Phys. Usp. {\bf 39}, 461 (1996)].
\bibitem{periodic} S. Khlebnikov, V. Rubakov, and P. Tinyakov, 
Nucl. Phys. B {\bf 367}, 334 (1991).
\bibitem{Bogoliubov} N. N. Bogoliubov, J. Phys. (Moscow) {\bf 11}, 23 (1947).
\bibitem{Magnus} P. Ao and D. J. Thouless, Phys. Rev. Lett. {\bf 70}, 2158
(1993); F. Gaitan, Phys. Rev. B {\bf 51}, 9061 (1995);
see also M. V. Feigel'man, V. B. Geshkenbein, A. I. Larkin, and 
V. M. Vinokur, Pis'ma ZhETF {\bf 62}, 811 (1995) 
[JETP Lett. {\bf 62}, 834 (1995)].
\bibitem{Zakharov&al} V. I. Zakharov, Nucl. Phys. B {\bf 371}, 637 (1992);
M. Porrati, Nucl. Phys. B {\bf 347}, 371 (1990).
\bibitem{Coleman} S. Coleman, Phys. Rev. D {\bf 15}, 2929 (1977); {\bf 16},
1248(E) (1977).
\bibitem{CC} C. G. Callan and S. Coleman, Phys. Rev. D {\bf 16}, 1762 (1977).
\bibitem{tables} A. Erd\'{e}lyi {\em et al.}, {\em Tables of Integral Transforms},
Vol. 1 (McGraw-Hill, New York, 1954), Eq. 1.3.7.
\bibitem{WG} U. Weiss and H. Grabert, Phys. Lett. A {\bf 108}, 63 (1985).
\bibitem{Matveev} V. V. Matveev, Phys. Lett. B {\bf 304}, 291 (1993) 
[hep-lat/9302006].
\bibitem{Vainstein&al} A. I. Vainstein, V. I. Zakharov, V. A. Novikov, 
and M. A. Shifman, Usp. Fiz. Nauk {\bf 136}, 553 (1982) 
[Sov. Phys. Usp. {\bf 25}, 195 (1982)].
\end{thebibliography}
\end{document}